\documentclass[pra,twocolumn,showpacs,superscriptaddress]{revtex4-1}

\usepackage{graphicx}
\usepackage{color}
\usepackage[usenames,dvipsnames]{xcolor}
\usepackage{amsmath,amsthm,amsfonts,amssymb,bm}

\newcommand{\nc}{\newcommand}
\nc{\eq}{\begin{equation}}
\nc{\eeq}{\end{equation}}
\nc{\eqa}{\begin{eqnarray}}
\nc{\eeqa}{\end{eqnarray}}
\nc{\ar}{\begin{array}}
\nc{\ear}{\end{array}}
\nc{\bfig}{\begin{figure}}
\nc{\efig}{\end{figure}}
\nc{\dg}{\dagger}
\nc{\eps}{\frac{\epsilon}{2}}
\nc{\juuri}{\sqrt{\Omega^2+(\eps)^2}}
\nc{\sx}{\sigma_x}
\nc{\sy}{\sigma_y}
\nc{\sz}{\sigma_z}
\nc{\spl}{\sigma_+}
\nc{\sm}{\sigma_-}
\nc{\Sx}{\bar{\sigma}_x}
\nc{\Sy}{\bar{\sigma}_y}
\nc{\Sz}{\bar{\sigma}_z}
\nc{\Spl}{\bar{\sigma}_+}
\nc{\Sm}{\bar{\sigma}_-}
\nc{\nn}{\nonumber}
\nc{\noi}{\noindent}
\nc{\omt}{\tilde{\omega}}
\nc{\Somt}{S(\omt)}
\nc{\Somtd}{S^{\dg}(\omt)}
\nc{\got}{\gamma_{\omega}(t)}
\nc{\gmot}{\gamma_{-\omega}(t)}
\nc{\po}{\mathcal{P}}
\nc{\qo}{\mathcal{Q}}
\nc{\adg}{a^{\dg}}
\nc{\gammat}{\tilde{\gamma}}
\nc{\Q}{$\mathcal{Q }$}
\nc{\C}{$\mathcal{C }$}
\nc{\kvec}{\mathbf{k}}

\def\bra#1{\mathinner{\langle{#1}|}}
\def\ket#1{\mathinner{|{#1}\rangle}}

\begin{document}

\title{Dynamical Memory Effects in Correlated Quantum Channels}

\author{Carole Addis}
\affiliation{SUPA, EPS/Physics, Heriot-Watt University, Edinburgh, EH14 4AS, UK}
\author{G\"{o}ktu\u{g} Karpat}
\affiliation{Faculdade de Ci\^encias, UNESP - Universidade Estadual Paulista, Bauru, SP, 17033-360, Brazil}
\author{Chiara Macchiavello}
\affiliation{Quit group, Dipartimento di Fisica, Universit� di Pavia, via A. Bassi 6, I-27100 Pavia, Italy}
\affiliation{Istituto Nazionale di Fisica Nucleare, Gruppo IV, via A. Bassi 6, I-27100 Pavia, Italy}
\author{Sabrina Maniscalco}
\email{smanis@utu.fi}
\affiliation{Turku Center for Quantum Physics, Department of Physics and Astronomy, University of Turku, FIN-20014, Turun yliopisto, Finland}
\affiliation{Centre for Quantum Engineering, Department of Applied Physics,  School of Science, Aalto University, P.O. Box 11000, FIN-00076 Aalto, Finland}

\date{\today}

\begin{abstract}
Memory effects play a fundamental role in the study of the dynamics of open quantum systems. There exist two conceptually distinct notions of memory discussed for quantum channels in the literature. In quantum information theory quantum channels with memory are characterised by the existence of correlations between successive applications of the channel on a sequence of quantum systems. In open quantum systems theory memory effects arise dynamically during the time evolution of quantum systems, and define non-Markovian dynamics. Here we relate and combine these two different concepts of memory. In particular, we study the interplay between correlations between multiple uses of quantum channels and non-Markovianity as non-divisibility of the $t$-parametrized family of channels defining the dynamical map.
\end{abstract}


\maketitle

\section{Introduction}

The theory of open quantum systems provides the necessary means to describe and analyse the interaction of a principal quantum system of interest with its surrounding environment \cite{book}. It is well know that the effects of this unavoidable interaction are in general detrimental for critical quantum traits present in the principal system, such as entanglement in composite systems. The study of open quantum systems has attracted considerable attention in recent years due to the fact that the preservation of genuine quantum properties, which serve as a resource for several different quantum information and communication protocols \cite{entreview,discordreview}, has become a very important challenge. In order to suppress the undesirable effects of environment induced decoherence, various methods have been put forward \cite{entnonmark,entdd,entzeno}. One such technique is through the exploitation of memory effects dynamically arising in the course of the time evolution of the system.

Memory effects emerge when an open quantum system interacts with its environment in a non-Markovian fashion. The characterization of non-Markovian quantum dynamics has been and still is a very significant problem in the study of open quantum systems \cite{nmreview1,nmreview2}. Numerous distinct criteria have been introduced to identify the non-Markovian memory effects, based on conceptually different approaches \cite{rivas10,hou11,lu10,luo12,bylicka14,chr14,fanchini14,breuer09}. Indeed, such memory effects have their roots at non-trivial temporal correlations among the states of the open system at different times throughout the dynamics. Besides, the emergence of memory effects is also known to be closely related to the dynamics of information exchange between the open system and its environment since future states of the system might depend on its past states when information flows back from the environment to the open system \cite{qjumps,breuer09,fanchini14,haseli14,bogna16}.

The concept of memory effects and non-Markovianity as information back-flow, which is typical of open quantum systems theory, does not however coincide with the concept of quantum channels with memory generally used in quantum information theory. The latter one, indeed, typically refers to the way a quantum channel (i.e., a quantum operation that is viewed as a channel to transfer information) acts on the system when it is used consecutively \cite{chiara,chanreview, memchan}. In particular, one indicates with memory or memoryless channels the situation in which multiple uses of the channel are correlated or independent from each other, respectively. In effect, the memory in this case is induced by the correlated action of noise channels on the system of interest consisting of a set of individual quantum systems, rather than the temporal correlations occurring throughout the time evolution of a single quantum system. To distinguish between these two different notions of memory, we shall use the term correlated  channels to describe the quantum channels with memory. On the other hand, the type of memory occurring due to the temporal correlations in the dynamics will be said to be non-Markovian memory effects.

Although both the non-Markovian memory effects and the memory due to the correlated application of quantum channels have been explored in the literature on their own as separate subjects, they have not yet been studied in relation to each other. In fact, our work aims to establish this link by investigating
the effect of classical correlations between multiple uses of quantum channels on the non-Markovian memory effects occurring as a result of the non-divisible nature of the dynamics. Specifically, considering a well established model for describing channels with memory \cite{chiara}, we examine how correlated application of quantum channels modify the non-Markovian memory effects, quantified via different measures of non-Markovianity, in a dephasing scenario.

This paper is organized as follows. In Sec. II., we introduce the type of open quantum system models that we intend to use in our study. In Sec. III, we discuss the identification and quantification of non-Markovian
memory effects for the considered model. In Sec. IV, we present the results of our investigation related to
the effect of correlated channels on the nature of non-Markovian dynamics. Sec. V includes the summary of our results.

\section{Correlated Quantum Channels}
 
Let us first introduce the type of classically correlated quantum channels that we consider in our
work. A single qubit Pauli channel, which is a random implementation of the Pauli transformations, is given by
\begin{equation} \label{pauli1}
\rho \rightarrow  \mathcal{E} (\rho)=\sum_{i=0}^3 q_i \sigma_i \rho \sigma_i,
\end{equation}
where the $q_i$'s constitute a probability distribution, i.e. $\sum_{i=0}^3q_i=1$, the $\sigma_0$ denotes the
$2\times2$ identity matrix and the $\sigma_i$'s are the Pauli operators in x,y,z directions. In the course of this
work, we focus our attention on two uses of quantum channels for the sake of simplicity. Provided that the noise is assumed to be uncorrelated for two uses of the channel, then the effect of such a channel can be described by independent applications of the considered map on the two-qubit state, that is
\begin{equation} \label{pauli2}
\rho \rightarrow  \mathcal{E} (\rho)=\sum_{i,j=0}^3 q_{i}q_{j} (\sigma_i \otimes \sigma_j)
\rho (\sigma_i \otimes \sigma_j)
\end{equation}
where $q_{i(j)}$ are independent probability distributions. 

However, it is possible to have some classical correlations in the repeated application of the channel which might modify the way the Pauli transformations act on the state, in which case we have
\begin{equation} \label{pauli2cor}
\rho \rightarrow  \mathcal{E} (\rho)=\sum_{i,j=0}^3 p_{ij} (\sigma_i \otimes \sigma_j) \rho
(\sigma_i \otimes \sigma_j),
\end{equation}
where $p_{ij}$ is not restricted to be factorized as $p_{ij}=q_iq_j$. A well studied model taking into account
the memory in the channel (in the form of classical correlations between multiple applications of the channel) has been proposed by Macchiavello and Palma \cite{chiara}, and its relevance has been discussed in the context of quantum information theory \cite{chanreview}. In this model, the joint probability distribution takes the the following form:
\begin{equation} \label{jointprob}
p_{ij}=(1-\mu)q_iq_j+\mu q_i\delta_{ij}.
\end{equation}
It is straightforward to observe that the above distribution implies the existence of an additional effect coming from the degree of classical correlation $\mu$, which with some probability forces the same Pauli transformation operator to be repeated in the second use of the channel.  When $\mu=0$ there are no correlations between the two uses of the channel.  On the contrary, the channel is fully correlated for $\mu=1$ and in this case it is guaranteed that the same Pauli transformation is applied on both qubits since the probability distribution is given by $p_{ij}=q_i\delta_{ij}$. 

In order to establish a link between the memory stemming from the correlated application of quantum channels and the non-Markovian memory effects due to the non-divisible dynamics, the coefficients $p_{ij}$ should explicitly depend on time. To this aim, we introduce a colored pure dephasing model describing the time evolution of a single qubit \cite{daffer04}, which admits a solution falling under the class of Pauli channels described by Eq. (\ref{pauli1}). This model allows us to explore the effect of the classical correlations, controlled by the parameter $\mu$, on the non-Markovian memory effects in the dynamics. 

Let us assume that the dynamics of a qubit is described by a time-dependent Hamiltonian
$H(t)=\hbar\Gamma(t)\sigma_z$, where $\Gamma(t)$ is an independent random variable with the statistics of a random telegraph signal. It can be written as $\Gamma(t)=\alpha n(t)$, where $n(t)$ has a Poisson distribution with a mean equal to $t/2\tau$ and $\alpha$ is a coin-flip random variable with the possible values $\pm \alpha$. If $\alpha=1$, the dynamics can be described by the following Kraus operators
\begin{align}
K_1(\nu) &= \sqrt{[1+\Phi(\nu)]/2}\mathbb{I}, \\
K_2(\nu) &= \sqrt{[1-\Phi(\nu)]/2}\sigma_3,
\end{align}
where we have $\Phi(\nu)=e^{-\nu}[\cos(u\nu)+\sin(u\nu)/u]$, and $u=\sqrt{(4\tau)^2-1}$
with $\nu=t/2\tau$ being the scaled time. Here, the parameter $\tau$ controls the degree of non-Markovianity of the dephasing process that produces the dynamical memory effects. Interested readers may refer to Ref. \cite{daffer04} for the technical details of the derivation and the solution of the model along with its physical relevance. 

For the above considered pure dephasing model, it is rather easy to verify that the time dependent
coefficients $q_i$'s in Eq. (\ref{pauli1}) take the form 
\begin{equation}
q_0 = \frac{1}{2}[1+\Phi(\nu)], \quad q_1=q_2=0, \quad q_3 = \frac{1}{2}[1-\Phi(\nu)].
\end{equation}
Hence, the correlated quantum channel in Eq. (\ref{pauli2cor}) now describes the dynamical evolution of the open system and it can be expressed in terms of the Kraus representation
\begin{align}  \label{dynamics}
\mathcal{E}(\rho)&=p_{03}(\sigma_0\otimes\sigma_3)\rho(\sigma_0\otimes\sigma_3) \nonumber \\
&+p_{30}(\sigma_3\otimes\sigma_0)\rho(\sigma_3\otimes\sigma_0)  \nonumber \\
&+p_{00}(\sigma_0\otimes\sigma_0)\rho(\sigma_0\otimes\sigma_0) \nonumber \\
&+p_{33}(\sigma_3\otimes\sigma_3)\rho(\sigma_3\otimes\sigma_3).
\end{align}
With this information at hand, we can study how the classical correlations quantified via the parameter
$\mu$ in the application of quantum channels affect the dynamically arising non-Markovian memory effects.

\section{Characterizing non-Markovian memory effects}
 
In this section, we will elaborate on the quantification of the non-Markovian memory effects in open quantum system dynamics. Despite the fact that there are many different ways of measuring the non-Markovian behavior of a quantum process \cite{nmreview1,nmreview2}, here we will mainly focus on two of them, which are relevant for our purposes.   

We commence by considering the well-known trace distance measure \cite{breuer09} (also known as the BLP measure) that is constructed upon the trace distance between two arbitrary states $\rho_1(t)$ and $\rho_2(t)$ of the system, given by
\begin{equation}
D(\rho_1(t),\rho_2(t))=\frac{1}{2} \rm{Tr}|\rho_1(t)-\rho_2(t)|,
\end{equation}
where $|A|=\sqrt{A^\dagger A}$. The trace distance measure has a physical interpretation in terms of the
distinguishability of two quantum states, variation of which during the evolution can be interpreted
as an information exchange between the principal system and its environment. Especially, a monotonic
loss of distinguishability between $\rho_1(t)$ and $\rho_2(t)$ throughout the dynamics, i.e. $dD(t)/dt<0$,
indicates that information flows from the system to the environment at all times. On the other hand, $dD(t)/dt>0$ implies that there exists a backflow of information from the environment back to the system, giving rise to a non-Markovian process. Based on this criterion, the BLP measure reads
\begin{align}
\mathcal{N}_{D}(\mathcal{E})&=&\max_{\rho_1(0),\rho_2(0)}\int_{(dD(t)/dt)>0}\frac{dD(t)}{dt}dt
\end{align}
where the maximum is taken over all possible pairs of initial states $\rho_1(0)$ and $\rho_2(0)$.
Markovian maps satisfy the property of divisibility, i.e., $\mathcal{E}_t=\mathcal{E}_{t,s}\mathcal{E}_s$ with $\mathcal{E}_{t,s}$ being CPTP and
$s \leq t$.
It is important to note that, although the trace distance is contractive (monotonically decreasing)
under completely positive and trace preserving (CPTP) maps, so that the distinguishability between $\rho_1(t)$ and $\rho_2(t)$ monotonically decreases for all divisible processes at all times, non-Markovianity based on trace distance is not exactly equivalent to the non-divisibility. Indeed, the BLP measure is only a witness for non-divisibility of quantum processes.
 
In addition, there exists a different class of non-Markovianity measures that exploit the fact that, entanglement, mutual information or some other information theoretic quantities are monotonically decreasing under local CPTP maps.  Differently from the case of BLP measure, here an ancillary system is introduced
with the same dimension as the principal system. Then, assuming that the map $\mathcal{E}_t$ acts only on the subsystem $B$ and the ancilla $A$ evolves trivially, the absence of dynamical memory effects immediately suggests that
\begin{equation}
X(( \mathbb{I} \otimes \mathcal{E}_t)\rho_{AB}) \leq X((\mathbb{I} \otimes \mathcal{E}_s)\rho_{AB}),
\label{ineq}
\end{equation}
at all times $0\leq s \leq t$ for all bipartite states $\rho_{AB}$, where $X$ is any considered monotonic quantity. Clearly, any violation of this inequality can be interpreted as a manifestation of non-Markovianity  since it signals that the intermediate map $\mathcal{E}_{t,s}$ is not a CPTP map, violating divisibility. In fact, it is straightforward to define a measure of memory effects, based on this violation by summing up the total increase of the chosen quantifier $X$ throughout the dynamics as
\begin{eqnarray} \label{Cmeasure}
\mathcal{N}(\mathcal{E})&=&\max_{\rho_{AB}}\int_{(dX(t)/dt)>0}\frac{dX(t)}{dt}dt
\end{eqnarray}
where the optimization should be performed over all bipartite states $\rho_{AB}$ in general. We should also mention in passing that such measures are nothing but witnesses for non-divisible dynamics, similarly to the case of the trace distance measure, although the BLP measure and these measures might lead to different conclusions in general.

Returning back to the correlated noise scenario, we recall that even in the simplest case a bipartite system
is required to analyse dynamical memory effects in correlated channels since we need two applications of the single qubit map on the system. Thus, studying the behavior of the quantity $X$ would require to consider a quadripartite 
state. This would make the optimization problem in the definition of the measures intractable in most situations.

Therefore, in order to be able to investigate the dynamical non-Markovian behavior due to non-divisibility in correlated noise scenario, we will slightly modify the characterization given in Eq. (\ref{ineq}). Specifically, we consider a principal system consisting of two qubits (as required to study the correlated quantum channels) without the addition of any ancillary qubits. We can then use the modified inequality
\begin{equation} \label{ineq2}
X(\mathcal{E}_t\rho_{AB}) \leq X(\mathcal{E}_s\rho_{AB}),
\end{equation}
at all times $0\leq s \leq t$ for all bipartite states $\rho_{AB}$. Note that the classically correlated quantum map in Eq. (\ref{dynamics}) no longer acts locally on the bipartite state. As a consequence, one cannot exploit the monotonicity property of certain quantities, such as mutual information, under local CPTP maps to detect the violation of divisibility. However, the type of correlated maps that we consider in Eq. (\ref{pauli2cor}) can be implemented by local operations and classical communication (LOCC), i.e., they belong to the class of LOCC maps since they are nothing but probabilistic unitary operations with local operators. For such quantum maps, we can utilize any entanglement measure as the quantifier $X$ to filter out the direct effect of classical correlations, for entanglement measures by definition are not only monotonic under local CPTP maps but also under local operations and classical communication. In our setting, if we assume that the intermediate map $\mathcal{E}_{t,s}$ cannot be a valid CPTP map unless it is LOCC, then the inequality in Eq. (\ref{ineq2}) is temporarily invalidated only for non-divisible dynamics.

\section{Non-Markovianity of classically correlated channels}
Having introduced both the type of correlated channels that we will use in our analysis and the non-Markovianity quantifiers, we can now study the effect of correlations in the channel on the non-Markovian dynamics. 
\begin{figure}[t]
\centering
\includegraphics[width=0.5\textwidth]{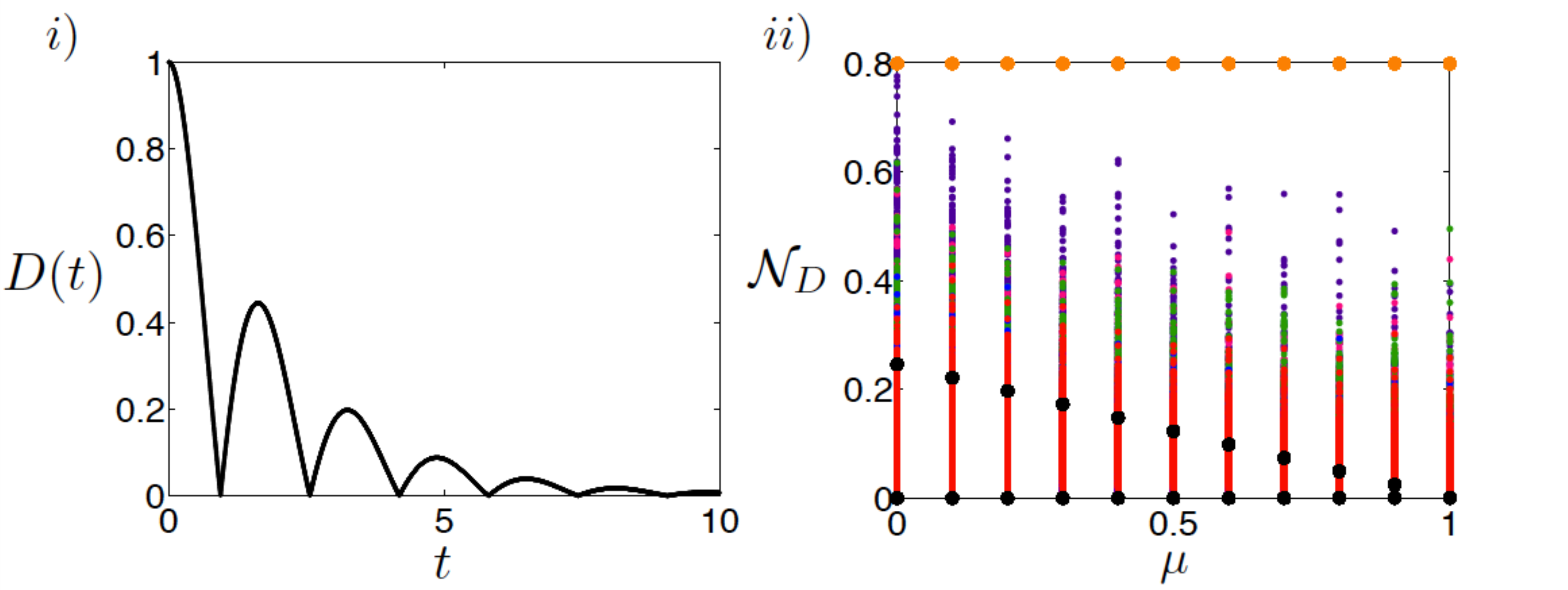}
\caption{a) Dynamics of the trace distance for the maximizing pair $\ket{\pm\pm}\bra{\pm\pm}$ for any value of the correlation parameter $\mu$., b) The BLP measure as a function of the correlation strength $\mu$ for $\tau=1$. The following random states are present: maximally entangled states (purple), pure states (pink), mixed states (red) and product states (green), combinations of mixed and pure states (blue), Bell states (black crosses) and the  product state $\ket{\pm\pm}\bra{\pm\pm}$ (orange stars). The optimal value is given by the highest point in the y-axis.}
\label{BLP}
\end{figure}

Using Eq. (\ref{dynamics}) we can write the time evolution of the density matrix of the system as follows
\begin{equation}
\rho(t) =
\rho(0)\circ\begin{pmatrix}
1&  \Phi(\nu) & \Phi(\nu) & \Gamma(\nu,\mu) \\
\Phi(\nu)& 1&\ \Gamma(\nu,\mu)& \Phi(\nu) \\
\Phi(\nu)   & \Gamma(\nu,\mu)  &1 &\Phi(\nu)  \\
\Gamma(\nu,\mu)& \Phi(\nu)& \Phi(\nu)& 1
\end{pmatrix}\nn
\end{equation}
where $\Gamma(\nu,\mu)= -\Phi(\nu)^2(-1+\mu)+\mu$. Note that only the anti-diagonal components depend on the correlation strength $\mu$. Using the analytical expression for the density matrix evolution we can calculate the BLP measure by numerical optimization over many pair of states. In Fig. \ref{BLP}(a) we display the dynamics of trace distance for the optimal pair, where the intervals of information backflow due to non-divisibility can be witnessed through the temporary increase of trace distance. In Fig. \ref{BLP}(b) we show how the BLP measure changes as we change the correlation coefficient $\mu$, sweeping between vanishing and full classical correlations for two consecutive uses of the channel. Note that in this plot the parameter $\tau$, controlling the degree of non-Markovianity of the map in absence of classical correlations, is fixed to 1 so that we can isolate the effect of the correlation parameter $\mu$ (non-Markovianity rises with increasing $\tau$). Performing an extensive numerical sampling of different pairs, we conclude that the optimal pair giving the maximum value of the BLP measure is always given by the separable states
\eq
\ket{\pm\pm}\bra{\pm\pm}=
\frac{1}{4}\begin{pmatrix}
1&\pm 1 & \pm1 & 1\\
\pm 1& 1& 1&\pm 1\\
\pm 1 & 1 & 1&\pm 1\\
1& \pm 1&\pm 1& 1
\end{pmatrix}
\eeq
In fact, from the plot one notices that the BLP measure seems to be 
independent of the effects of correlations in the application of the channel. 
A closer inspection reveals that the trace distance for the pair of optimal states reads
\begin{equation}
D=\frac{1}{2}\sum|\Lambda_i| =|\Phi(\nu)|,
\end{equation}
where $\Lambda_i$ are the eigenvalues of $\rho_1(t)-\rho_2(t)=\rho_{12}(t)$. Consequently, non-Markovianity as measured by trace distance is completely insensitive to classical correlations in multiple applications of the channel. Stated another way, dynamical non-Markovianity does not depend on whether or not the channel here has correlations.

\begin{figure}[t]
\centering
\includegraphics[width=0.48\textwidth]{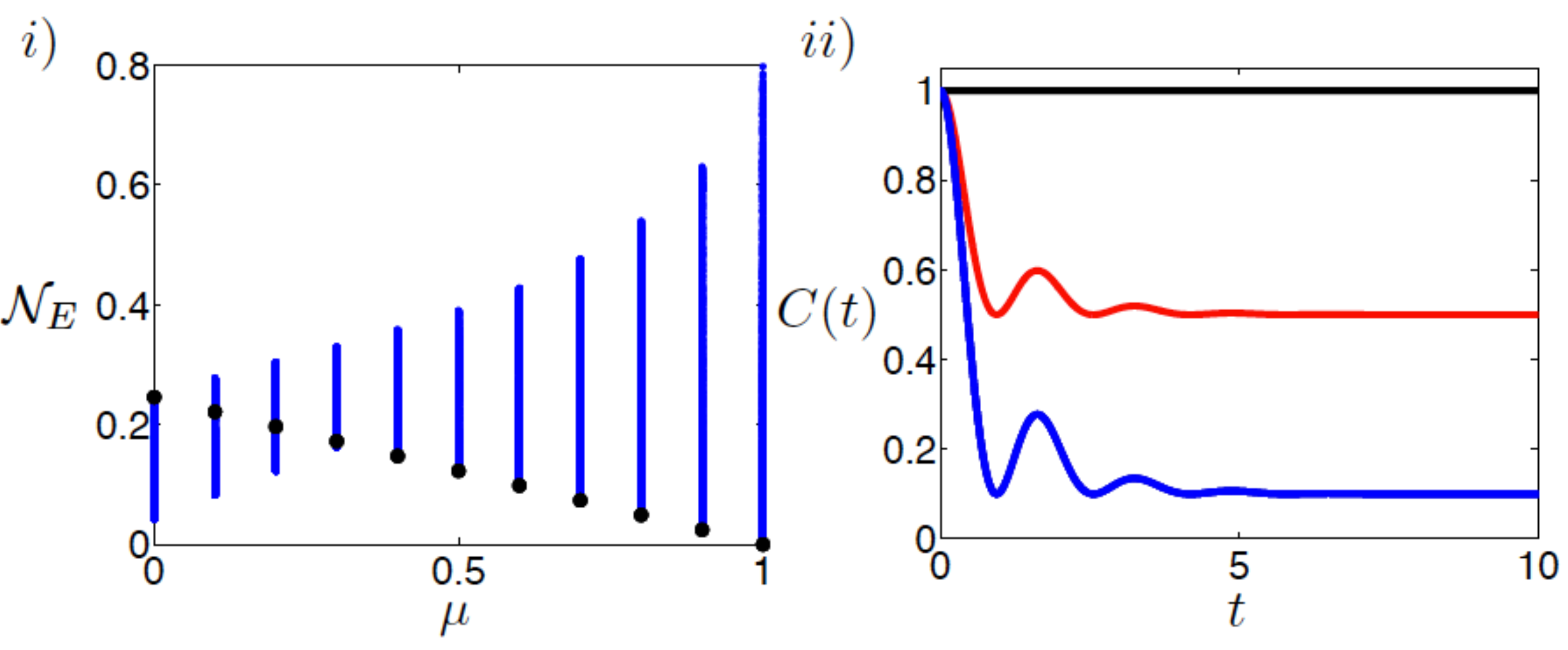}
\caption{The entanglement based measure for $\tau=1$. a) We plot random maximally entangled states (blue), implemented by applying local unitaries to Bell states, and the Bell states (black crosses). In b), we plot the concurrence of the Bell states for $\mu=1$ (solid black), $\mu=0.5$ (blue dashed) and $\mu=0.1$ (red dot dashed).}\label{C}
\end{figure}

Let us now turn our attention to the entanglement based measure of non-Markovian memory effects described in Eq. (\ref{Cmeasure}). We choose $X$ to be the 
concurrence, which is defined as
\begin{equation}
C(\rho)=\text{max}(0,\sqrt{\lambda_1}-\sqrt{\lambda_2}-\sqrt{\lambda_3}-\sqrt{\lambda_4}),
\end{equation}
with \{$\lambda_i$\}  the eigenvalues of the Hermitian operator $R=\rho(\sigma_y\otimes\sigma_y)\rho^*(\sigma_y\otimes\sigma_y)$ in decreasing order and  $\rho^*$ the complex conjugate of the density matrix $\rho$. In Fig. \ref{C}(a) we present the results of our analysis for the entanglement based measure of non-Markovianity performing a numerical sampling of many different types of initial bipartite states. For convenience, we only display the Bell states (black points) and the family of maximally entangled states obtained from the Bell states by applying local unitary operations (blue points), even though our sampling included many more different types of initial states. Unlike the BLP measure, the maximizing state and the optimal value of the entanglement based measure are clearly dependent on the correlation parameter $\mu$ in this case. Particularly, non-Markovian memory effects are strengthened as the amount of classical correlations in the channel are increased. This demonstrates a fundamental difference between these two approaches to the quantification of dynamical memory effects when they are analysed in relation with the classical correlations in the 
operation of quantum channels.

For Bell states the concurrence can be written as
\eq
C(t)=-\Phi(\nu)^2(-1+\mu)+\mu.
\eeq
Hence for $\mu=1$, in the case of fully correlated channels, entanglement is frozen at unity as can be seen in Fig. \ref{C}(b). Note that the diagonal elements of the density matrix are constant for the pure dephasing case. Moreover, when the channels are fully correlated ($\mu=1$), the anti-diagonal elements are also constant in time. Hence, any X-shaped state including the Bell states do not evolve in time for fully correlated channels. As the correlation parameter $\mu$ decreases, the time-dependency of the Bell states becomes more and more dominant and the concurrence begins to decay increasingly. On the other hand, the amount of revivals and thus non-Markovian behavior gets amplified as well. Looking at the figure, we also see that the Bell states maximize the measure only for $\mu=0$ (in case of uncorrelated channels). For $\mu>0$, other maximally entangled states, that can be obtained from the Bell states by local application of unitary operations, optimize the measure. It is worth noticing that if one would assume that the Bell states are the optimal ones, as it has been usually done in the literature, one could get a completely wrong idea about the non-Markovian behavior of the dynamics.

As a final remark we emphasize that the intervals of temporary revivals for both trace distance and concurrence fully coincide (even though this is not immediately obvious by comparing Fig. \ref{BLP}(a) and Fig. \ref{C}(b) since revivals in concurrence become very small as time passes). Since the revivals in the trace distance always imply that the intermediate map $\mathcal{E}_{t,s}$ is not CPTP during these time intervals, one can conclude that the revivals in entanglement are also due to the non-divisibility of the map rather than to the fact that $\mathcal{E}_{t,s}$ cannot be implemented by LOCC operations. Note that, despite the fact that we cannot prove in general the validity of our assumption, namely that the intermediate map $\mathcal{E}_{t,s}$ is non-CPTP unless it is LOCC, we observe that it can be justified in the studied example. Consequently, both the trace distance measure and the entanglement based measure quantify the revivals occurring purely as a result of non-divisibility properties of the map. Nonetheless, we should always keep in mind that both these quantifiers are just witnesses for non-divisible dynamics. Indeed, we have seen that while entanglement based measure can detect the effects of the classical correlations in the channel, trace distance based measure does not feel such effects even for the fully correlated case, which points out to a remarkable difference between the two. 

\vspace{-0.2cm}

\section{Conclusion} 

In summary, we have explored the effect of correlations in the quantum channels with correlated noise on the dynamical memory effects stemming from the non-Markovian dynamics of the open quantum system. Particularly, under a well motivated model for quantum channels with memory, we have investigated the role of correlations between uses of quantum channels in modifying the non-Markovian memory effects arising throughout the time evolution of the system. 

For this purpose, we have considered a colored pure dephasing model with non-Markovian characteristics. Our analysis has unveiled that the classical correlations present in the studied quantum channels do not affect the non-Markovian features of the dynamics in any way, when we quantify the memory effects through the trace distance measure. On the other hand, we have demonstrated that, if we choose to utilize the entanglement based measure, correlations between the multiple applications of the quantum channels can indeed amplify the dynamical non-Markovian memory effects. Therefore, our investigation reveals a clear difference between these two widely used measures of non-Markovianity.

We should finally mention that even though we have examined a particular model, which describes the correlations between the consecutive uses of quantum channels, and also considered a specific type of dephasing dynamics, our treatment can be easily applied to study more general scenarios in a straightforward way.

\vspace{-0.1cm}

\begin{acknowledgements}
C.A. acknowledges financial support from the EPSRC (UK) via the Doctoral Training Centre in Condensed Matter Physics under grant number EP/G03673X/1. G. K. is grateful to Sao Paulo Research Foundation (FAPESP) for the fellowship given under grant number 2012/18558-5. S. M. acknowledges financial support from the EU Collaborative project QuProCS (Grant Agreement 641277), the Academy of Finland (Project no. 287750), and the Magnus Ehrnrooth Foundation.
\end{acknowledgements}

\end{document}